\documentclass{emulateapj}
\usepackage{txfonts}

\shorttitle{Properties of M31 globular clusters}

\shortauthors{Wang et al.}

\begin{document}
\slugcomment{AJ, in press}
\title{Spectral energy distributions and age estimates of 104 M31 globular clusters}

\author{Song Wang,\altaffilmark{1,2,3} Zhou Fan,\altaffilmark{1} Jun
  Ma,\altaffilmark{1,3} Richard de Grijs,\altaffilmark{4,5} Xu
  Zhou\altaffilmark{1}}

\altaffiltext{1}{National Astronomical Observatories, Chinese Academy
  of Sciences, Beijing, 100012, P. R. China;\\ majun@bac.pku.edu.cn}

\altaffiltext{2}{Graduate University, Chinese Academy of Sciences,
  Beijing, 100039, P. R. China}

\altaffiltext{3}{Key Laboratory of Optical Astronomy, National Astronomical Observatories,
 Chinese Academy of Sciences, Beijing, 100012, China}

\altaffiltext{4}{Kavli Institute for Astronomy and Astrophysics,
  Peking University, Beijing, 100871, P. R. China}

\altaffiltext{5}{Department of Physics \& Astronomy, University of
  Sheffield, Sheffield S3 7RH, UK}

\begin{abstract}
We present photometry of 104 M31 globular clusters (GCs) and GC candidates in 15 intermediate-band
filters of the Beijing-Arizona-Taiwan-Connecticut (BATC) photometric system. The GCs and GC
candidates were selected from the Revised Bologna Catalog (v.3.5). We obtain the cluster ages by
comparing the photometric data with up-to-date theoretical synthesis models. The photometric
data used are {\sl GALEX} far- and near-ultraviolet and 2MASS near-infrared $JHK_{\rm s}$
magnitudes, combined with optical photometry. The ages of our sample clusters cover a large range,
although most clusters are younger than 10 Gyr. Combined with the ages obtained in our series of
previous papers focusing on the M31 GC system, we present the full M31 GC age distribution. The
M31 GC system contains populations of young and intermediate-age GCs, as well as the `usual'
complement of well-known old GCs, i.e., GCs of similar age as the majority of the Galactic GCs. In
addition, young GCs (and GC candidates) are distributed nearly uniformly in radial distance from
the center of M31, while most old GCs (and GC candidates) are more strongly concentrated.
\end{abstract}


\section{Introduction}
\label{Introduction.sec}

Globular clusters (GCs) are among the oldest known stellar systems in the Universe. They
typically have ages similar to those of their host galaxies, thus making them fossils that may
provide important information about the formation and evolution of their parent galaxies. In
addition, nearly all types of galaxies contain GCs, from dwarfs to giants and from the earliest to
the latest types \citep{Fusi05}. However, our most in-depth understanding of GC systems has
predominantly come from studies of the Milky Way.

M31, located at a distance of $\sim 780$ kpc \citep{sg98, mac01}, is the largest galaxy in the
Local Group. By virtue of the natural advantage of being located at a reasonable distance, the
galaxy offers us an ideal environment for detailed, resolved investigations of a large GC system,
using both {\sl Hubble Space Telescope (HST)} \citep[e.g.,][]{grill96,hfr97,rich05,perina09b} and
ground-based observations with large telescopes \citep[e.g.,][]{ch91}.

A large number of studies focusing on the M31 GC system have been performed since
\citet{hubble32}'s original identification of 140 GC candidates in M31. The latest Revised Bologna
Catalogue of M31 GCs and candidates (hereafter RBC v.3.5) \citep{gall04,gall06,gall07} was updated
on March 27, 2008, and contains 1983 objects (509 confirmed and 1049 candidate GCs, 9
controversial objects, 147 galaxies, 6 H{\sc ii} regions, 245 stars, 5 asterisms, and 13 extended
clusters). These objects were observed and discovered by a large number of authors using a variety
of observational systems \citep[see, e.g.,][]{vete62a, sarg77, batt80, Crampton85, bh00}. To
obtain a homogeneous photometric data set, \citet{gall04} took the observed data of \citet{bh00}
as reference and transformed other observations to this standard setup.

An accurate and reliable analysis of star clusters is important for our understanding of the
formation, buildup, and evolutionary processes in galaxies. By comparing integrated photometry
with models of simple stellar populations (SSPs), recent studies have achieved some success in
determining ages and masses of extragalactic star clusters
\citep[e.g.,][]{degrijs03a,degrijs03b,degrijs03c,degrijs06,bik03,ma06,fan06,ma07a,ma09}.
\citet{ma06} and \citet{fan06} derived age estimates for M31 GCs by fitting SSP models
\citep[][henceforth BC03]{bru03} to their photometric measurements in a large number of
intermediate- and broad-band filters spanning the spectral range from the optical to the
near-infrared (NIR). In particular, \citet{ma07a} determined an age for the M31 GC S312 (B379),
using multicolor photometry from the near-ultraviolet (near-UV) to the NIR, of
$9.5^{+1.2}_{-1.0}$~Gyr. S312 (B379) is, in fact, among the first extragalactic GCs for which the
age was estimated accurately and independently, using main-sequence photometry, at
$10^{+2.5}_{-1}$ Gyr \citep{brown04}. This provides a robust check on our methodology to derive
age constraints based on the spectral energy distributions (SEDs) of (simple) stellar systems.

This paper is organized as follows. In \S\ref{data.sec} we present
Beijing-Arizona-Taiwan-Connecticut (BATC) observations of our sample GCs and GC candidates, the
relevant data-processing steps, and the {\sl GALEX} (far- and near-UV), optical broad-band, and
Two-Micron All Sky Survey (2MASS) NIR data that are subsequently used in our analysis. In
\S\ref{fit.sec} we derive the ages of our sample clusters by comparing their SEDs with the {\sc
galev} SSP models. We then discuss and summarize our results in \S\ref{result.sec}.

\section{GC sample and BATC intermediate-band photometry}
\label{data.sec}

\subsection{GC sample selection}

To obtain photometry in 15 intermediate-band filters of the BATC photometric system for 61 GCs and
GC candidates in the RBC v.3.5, for which few measurements are presently available in any
photometric system, \citet{fan09} mined the BATC survey archive for observations obtained between
February 1995 and March 2008. The resulting set of observations covers approximately six square
degrees. For the purpose of estimating accurate cluster ages, we selected clusters for which the
metallicities and reddening values had been estimated accurately, independently, and homogeneously
in previous studies \citep{hbk91,bh00,per02,fan08}: see \S2.6. We selected classes 1, 2, 3, and 8
(1580 objects) from column `f' in the RBC v.3.5, which include GCs, candidate GCs, controversial
objects, and extended clusters. This resulted in an initial selection of 366 objects.
\citet{jiang03}, \citet{ma06}, and \citet{ma09} obtained multicolor photometry for 180 of these
GCs and GC candidates. In this paper, we consider the remaining 186 objects. \citet{caldwell09}
published an updated catalog of 1300 objects in M31, including 670 likely star clusters, with the
remaining objects being stars or background galaxies once thought to be clusters \citep[see Tables
3 and 5 of][]{caldwell09}. From a comparison with \citet{caldwell09}, we find that 66 objects are
either stars or background galaxies. Therefore, the final sample of M31 GCs and GC candidates
analyzed in this paper includes 120 objects. However, we cannot obtain accurate photometric
measurements for 16 of these objects because of either a nearby very bright object (B065 and
B344D), very faint fluxes superimposed onto a bright background (B119, B396, NB16, and V031), or a
location very close to (or blend with) another object (B150D, B176, B256D, B302, B345, B366, B381,
B391, and B397), leading to compromised photometric measurements. Object B330 is both faint and
located very close to a brighter object. Thus, here we analyze the multicolor photometric
properties of 104 GCs and GC candidates. Figure 1 shows their spatial distribution across the M31
fields observed with the BATC multicolor system.

\begin{figure}
\resizebox{\hsize}{!}{\rotatebox{0}{\includegraphics{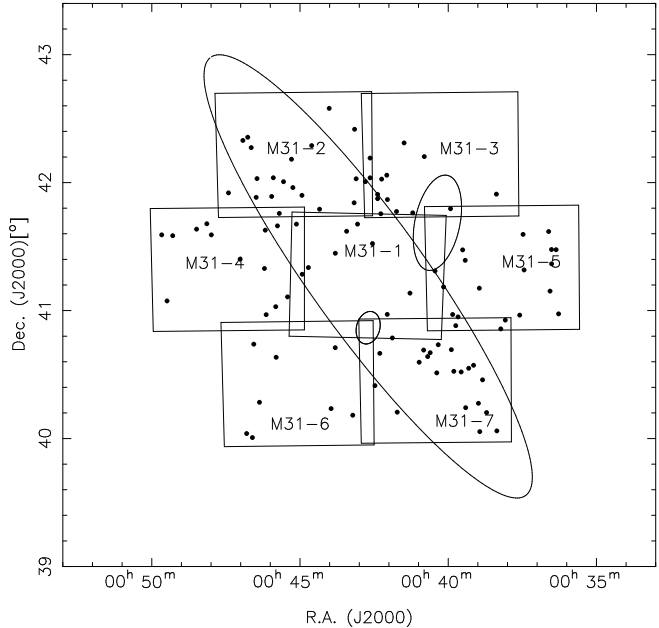}}}

\caption{BATC observations of our M31 fields. Each field is $58 \arcmin\times 58 \arcmin$ (size of
the old CCD). The large ellipse is the boundary between the M31 disk and halo \citep{rac91}, while
the two small ellipses represent the $D_{25}$ isophotes of NGC 205 (northwest) and M32
(southeast). Solid circles indicate the sample GCs and GC candidates discussed in this paper.}
\label{fig1}
\end{figure}

\subsection{BATC intermediate-band photometry}

The observations of our sample GCs and GC candidates were carried out in the BATC photometric
system, using the 60/90 cm $f$/3 Schmidt telescope at Xinglong Station of the National
Astronomical Observatories of the Chinese Academy Sciences (NAOC). The BATC system includes 15
intermediate-band filters, covering a wavelength range from 3300 {\AA} to 1 $\mu$m. The parameters
of the filters are given in Table 1, where column (1) gives the filter name, column (2) is the
central wavelength for each filter, and column (3) lists the bandwidth for each filter. The
2k$\times$2k CCD used before February 2006 had a pixel size of 15 $\mu$m and a resolution of
$1.7''$ pixel$^{-1}$. After February 2006, a new 4k$\times$4k CCD with a pixel size of 12 $\mu$m
was used, with a resolution of $1.3''$ pixel$^{-1}$ \citep{fan09}. The new CCD camera is much more
sensitive at short wavelengths.

We obtained 143.9 hours of imaging (447 images) of the M31 field, covering about six square
degrees, through the set of 15 filters in five observing runs from 1995 to 2008, spanning 13 years
\citep[see for details][]{fan09}. The data were reduced using standard procedures, including bias
subtraction and flat fielding of the CCD images, with an automatic data-reduction software package
(PIPELINE I) specifically developed for the BATC sky survey. BATC magnitudes are defined and
obtained in a similar way as for the spectrophotometric AB magnitude system \citep[see for
details][]{ma09}. For the $a$ to $p$ filters of the central field of M31 (M31-1 in Figure 1), the
absolute flux of the combined images was obtained using calibrated standard stars, while for the
M31-2 to M31-7 fields we used the M31-1 field to derive secondary transformations \citep[see for
details][]{fan09}.

We determined the magnitudes of our sample objects on the combined images using standard aperture
photometry, i.e., using the PHOT routine in DAOPHOT \citep{stet87}. To avoid contamination from
nearby objects, we adopted apertures with radii of 3 and 4 pixels on the 2k$\times$2k and
4k$\times$4k CCDs, respectively. For the old CCD, we took 8 and 13 pixels from the object's center
as the inner and outer radii of the sky annulus for background determination, while for the new
CCD, the corresponding radii were set at 10 and 17 pixels, respectively \citep{fan09}. We used
isolated stars to obtain point-source aperture corrections by measuring the magnitude differences
between the fluxes contained within radii of 3 (4) pixels on the old (new) CCD images and the
total stellar magnitudes in each of the 15 BATC filters. The resulting aperture-corrected SEDs for
the sample GCs and GC candidates in M31 are provided in Table 2. Columns (2) to (16) represent the
magnitudes in the 15 BATC passbands used for our photometry. The $1\sigma$ magnitude
uncertainties, from DAOPHOT, are listed for each object on the second line for the corresponding
passband. For some GCs and GC candidates, the magnitudes in some filters could not obtained
because of low signal-to-noise ratios.

\subsection{{\sl GALEX} UV, optical broad-band, and 2MASS NIR photometry}

To estimate the ages of the M31 GCs and GC candidates, we should ideally use as many photometric
data points covering as wide a wavelength range as possible
\citep[cf.][]{degrijs03b,Anders04,ma09}. The RBC v.3.5 includes {\sl GALEX} (far- and near-UV)
fluxes from \citet{rey07}, optical broad-band, and 2MASS NIR magnitudes for 1983 objects, which we
use as the basis for our analysis. Although the $UBVRI$ magnitudes of the objects published by
\citet{bh00} are included in the RBC v.3.5 and as such provide the most homogeneous set of
photometric measurements available, the relevant photometric uncertainties are not listed.
Therefore, we adopt the original $UBVRI$ measurements of \citet{bh00}, including their published
photometric errors. For the remaining objects we adopt the $UBVRI$ measurements from the RBC
v.3.5. We assign photometric uncertainties following \citet{gall04}, i.e., $\pm0.05$ mag in $BVRI$
and $\pm0.08$ mag in $U$ \citep[see for details][]{ma09}.

In the RBC v.3.5, the 2MASS $JHK_{\rm s}$ magnitudes were transformed to the CIT photometric
system \citep{gall04}. However, we need the original 2MASS $JHK_{\rm s}$ data to compare the
observed SEDs with the SSP models, so we reversed the transformation using the equations given by
\citet{Carpenter01}. We obtained the magnitude errors in the $JHK_{\rm s}$ bands by comparing our
photometric data with fig. 2 of \citet{Carpenteretal01}, which shows the generic photometric
uncertainties as a function of magnitude for stars brighter than their observational completeness
limits \citep[see for details][]{ma09}. We include the {\sl GALEX}, optical broad-band, and 2MASS
NIR photometry of the sample clusters in Table 3 (columns 3 to 12), where the photometric errors
are listed for each object on the second line for the corresponding passband. The {\sl GALEX}
photometric system is calibrated to match the spectrophotometric AB system, while the optical
broad-band and 2MASS photometric data are given in Vega magnitudes. Finally, column 2 includes the
classification flags from the RBC v.3.5.

\subsection{Comparison with previously published photometry}

To check our photometry, we transformed the BATC intermediate-band system to the $UBVRI$
broad-band system using the relationships between these two systems derived by \citet{zhou03}:

\begin{equation}
B=m_{d}+0.2201(m_{c}-m_{e})+0.1278\pm0.076 \quad \mbox{and}
\end{equation}

\begin{equation}
V=m_{g}+0.3292(m_{f}-m_{h})+0.0476\pm0.027.
\end{equation}

$B$-band photometry can be derived from the BATC $c, d$, and $e$ bands, while $V$-band magnitudes
can be obtained from the BATC $f, g$, and $h$ bands. Figure 2 shows a comparison of the $B$ and
$V$ photometry of our M31 sample objects with previous measurements from \citet{bh00} and
\citet{gall04}. The mean $B$ and $V$ magnitude differences---in the sense of this paper minus
\citet{bh00} or \citet{gall04}---are $\langle \Delta B \rangle =-0.077\pm 0.022$ mag and $\langle
\Delta V \rangle =-0.047\pm0.033$. Our magnitudes are in good agreement with previous $V$-band
determinations. However, a significant disagreement becomes apparent in the $B$ band for
objects with $B > 17.5$ mag. This disagreement has its origin in the difference between our
photometry and that of \citet{gall04}. In fact, our $B$-band photometry agrees well with that of
\citet{bh00}, even for $B > 17.5$ mag objects (except for one sample cluster). Referring to
\citet{jiang03} and \citet{ma09}, we also see that our photometric values are fully consistent
with \citet{bh00}. In \citet{ma09}, the analysis of the majority of the GCs was based on the
photometric data of \citet{bh00}, so even in the $B$ band the agreement is good: see fig. 3 of
\citet{ma09}. We excluded B257 from the comparison, because its $V$-band magnitude is too faint
compared with the magnitudes obtained in the other bands (see Table 3). In fact, from the observed
SEDs, this photometric measurement is unusually far away from the best-fitting integrated SEDs
(see \S3 for more details). This data point was taken from Table 4 of \citet{bh01}, and the offset
may be a typical error. Based on the BATC $f, g$, and $h$ magnitudes, we obtain $V = 17.76$ mag
for B257. Note that the $B$-band magnitude of this cluster as listed in Table 4 of \citet{bh01} is
11.907, which is too bright for any reasonable SED and may also be a typical uncertainty.

\begin{figure}
\resizebox{\hsize}{!}{\rotatebox{-90}{\includegraphics{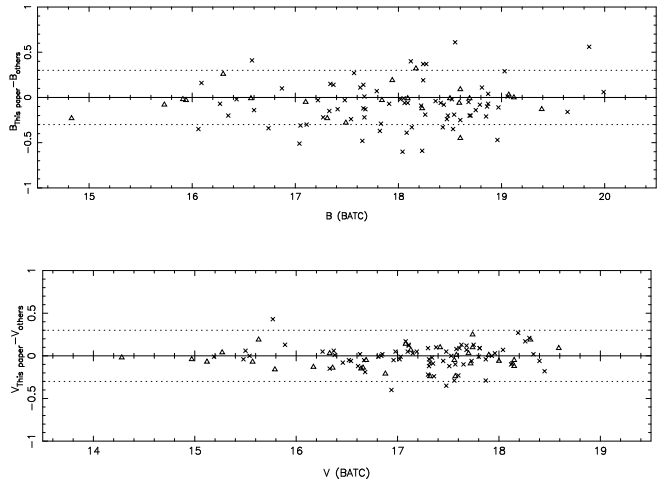}}}

\caption{Comparison of the photometry of our GCs and GC candidates with previous measurements from
\citet{bh00} (triangles) and \citet{gall04} (crosses). The dashed lines enclose $\Delta V$ and
$\Delta B = 0.3$ mag.} \label{fig2}
\end{figure}

\subsection{Metallicities and reddening values}

We require independently (but homogeneously) determined metallicities and reddening values to
robustly and accurately estimate the ages of our sample objects.  We used three homogeneous
sources of spectroscopic metallicities \citep{hbk91, bh00, per02} and one homogenized reference
\citep[see for details][]{fan08, ma09}.

Following \citet{ma09}, for reasons of consistency we ranked the sources used to assign
metallicities to our M31 GCs in order of preference. Metallicities from \citet{per02} were chosen
whenever available because of the large number of their metallicity determinations, followed (in
order) by metallicity determinations from \citet{bh00} and \citet{hbk91}. If none of these three
sources included spectroscopic metallicities for a given sample cluster, we used the corresponding
value from \citet{fan08}.

For reddening values, we used \citet{bh00} and \citet{fan08} as our reference \citep[see for
details][]{ma09}. Because the reddening values from \citet{fan08} comprise a homogeneous data set
and the number of GCs included is greater than that of \citet{bh00}, we preferentially adopt
\citet{fan08} reddening values, followed by those of \citet{bh00}, in a similar approach as
adopted by \citet{ma09}. The metallicities and reddening values adopted for our sample clusters
are listed in Table 4.

\section{Age determination}
\label{fit.sec}

An SSP is defined as a single generation of coeval stars characterized by the same parameters,
including metallicity, age, and stellar initial mass function (IMF). SSP models are calculated on
the basis of a set of evolutionary tracks of stars of different initial masses, combined with
stellar spectra at different evolutionary stages. In this paper, and following \citet{ma09}, we
compare the SEDs of our sample objects with the {\sc galev} SSP models
\citep[e.g.,][]{kurth99,schulz02,anders03} to estimate their ages. The {\sc galev} SSPs are based
on the Padova isochrones (covering wavelengths from 91 {\AA} to 160 $\mu$m) and a \citet{sal55}
stellar IMF with lower and upper mass limits of $0.10~{\rm M}_\odot$ and 50--70 ${\rm M}_\odot$,
respectively (the latter depending on metallicity). These models cover ages from 4 Myr to 16 Gyr,
with an age resolution of 4 Myr for ages younger than 2.35 Gyr, and 20 Myr for older ages. We
convolved the theoretical SSP SEDs with the {\sl GALEX}, broad-band $UBVRI$, BATC
intermediate-band, and 2MASS $JHK_{\rm s}$ filter response curves to obtain synthetic UV, optical,
and NIR photometry \citep{ma09}. The synthetic magnitude in the AB magnitude system for the
$i^{\rm th}$ filter is

\begin{equation}
m_i=-2.5\log\frac{\int_{\nu}F_{\nu}\varphi_{i} (\nu){\rm d}\nu}{\int_{\nu}\varphi_{i}(\nu){\rm
d}\nu}-48.60,
\end{equation}
where $F_{\nu}$ is the theoretical SSP SED (which is a function of age and metallicity) and
$\varphi_i$ is the response curve of the $i^{\rm th}$ filter. The {\sc galev} SSP models include
five initial metallicities, $Z=0.0004, 0.004, 0.008, 0.02$ (solar), and $0.05$. For other
metallicities, the relevant spectra can be obtained by linear interpolation of the appropriate
model spectra for any of these five metallicities. For metallicities below $Z=0.0004$ we use the
$Z=0.0004$ model \citep{ma09}.

To determine the most compatible {\sc galev} SSP model for a given observed SED, we adopted a
$\chi^2$ minimization test,

\begin{equation}
\chi^2=\sum_{i=1}^{25}{\frac{[m_{\nu_i}^{\rm intr}-m_{\nu_i}^{\rm
mod}(t)]^2}{\sigma_{i}^{2}}},
\end{equation}
where $m_{\nu_i}^{\rm mod}(t)$ is the magnitude in the $i^{\rm th}$ filter of a theoretical SSP at
age $t$, while $m_{\nu_i}^{\rm intr}$ is the intrinsic (observed and corrected) magnitude in the
same filter. The interstellar extinction curve $A_{\nu}$ is taken from \citet{car89},
$R_{\nu}=A_V/E_{B-V}=3.1$. $\sigma_{i}$ is the magnitude uncertainty in the $i^{\rm th}$ filter,
defined as

\begin{equation}
\sigma_i^{2}=\sigma_{{\rm obs},i}^{2}+\sigma_{{\rm mod},i}^{2}.
\end{equation}
Here, $\sigma_{{\rm obs},i}$ and $\sigma_{{\rm mod},i}$ are the observational uncertainty and that
associated with the model itself, respectively. \citet{charlot96} estimated $\sigma_{{\rm mod},i}$
by comparing the colors obtained from different stellar evolutionary tracks and spectral
libraries. We adopt $\sigma_{{\rm mod},i}=0.05$ mag, following \citet{wu05}, \citet{ma06, ma09},
and \citet{fan06}. The SED fits of our sample GCs and GC candidates are shown in Fig. 3.

\begin{figure*}
\vspace{-1.0cm}
\figurenum{3} \resizebox{\hsize}{!}{\rotatebox{-0}{\includegraphics{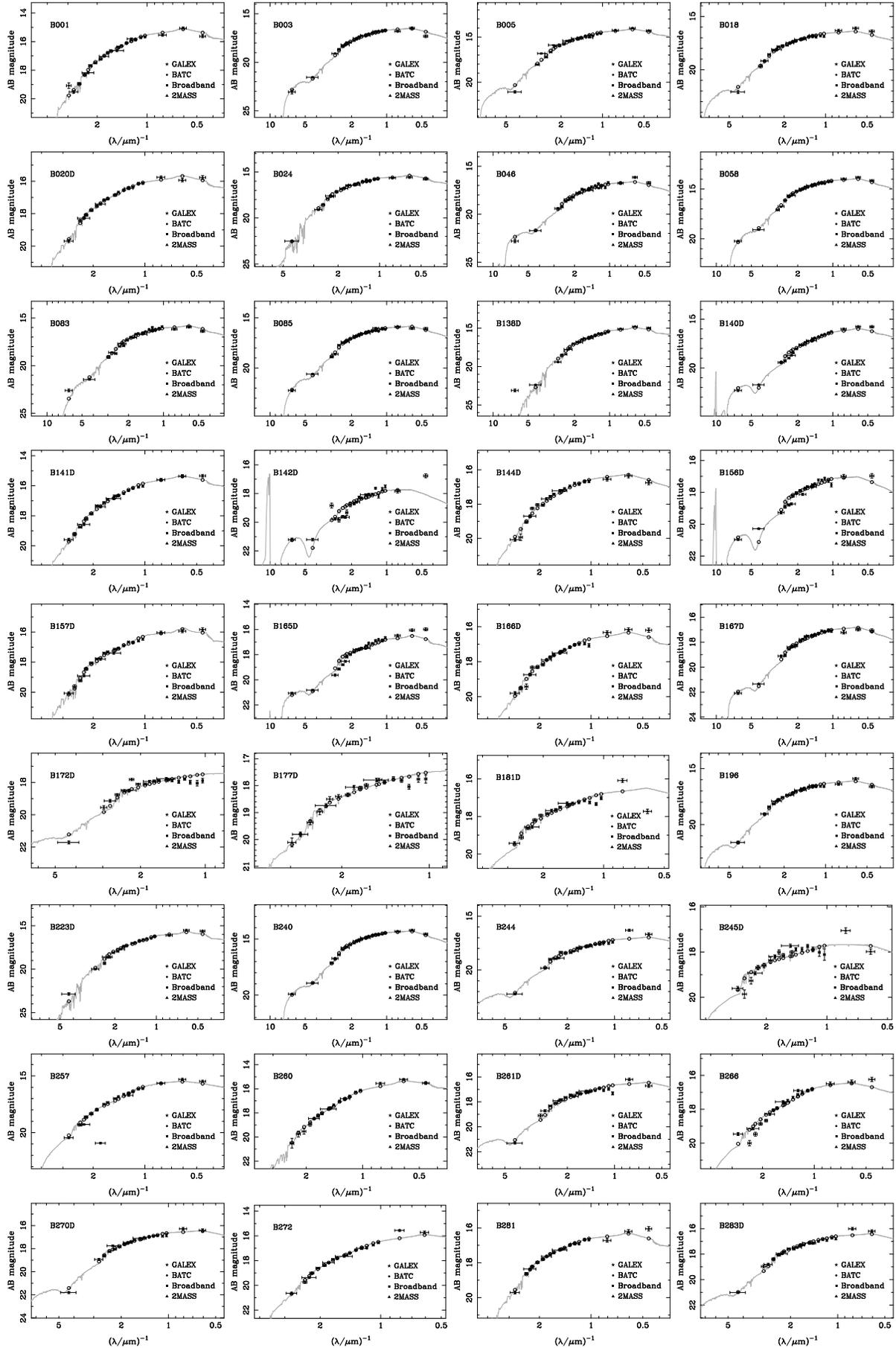}}}
\vspace{0.2cm} 
\caption{SED fits of the {\sc galev} SSP models to our sample objects.}
\label{fig3}
\end{figure*}

\begin{figure*}
\vspace{-1.0cm}
\figurenum{3} \resizebox{\hsize}{!}{\rotatebox{-0}{\includegraphics{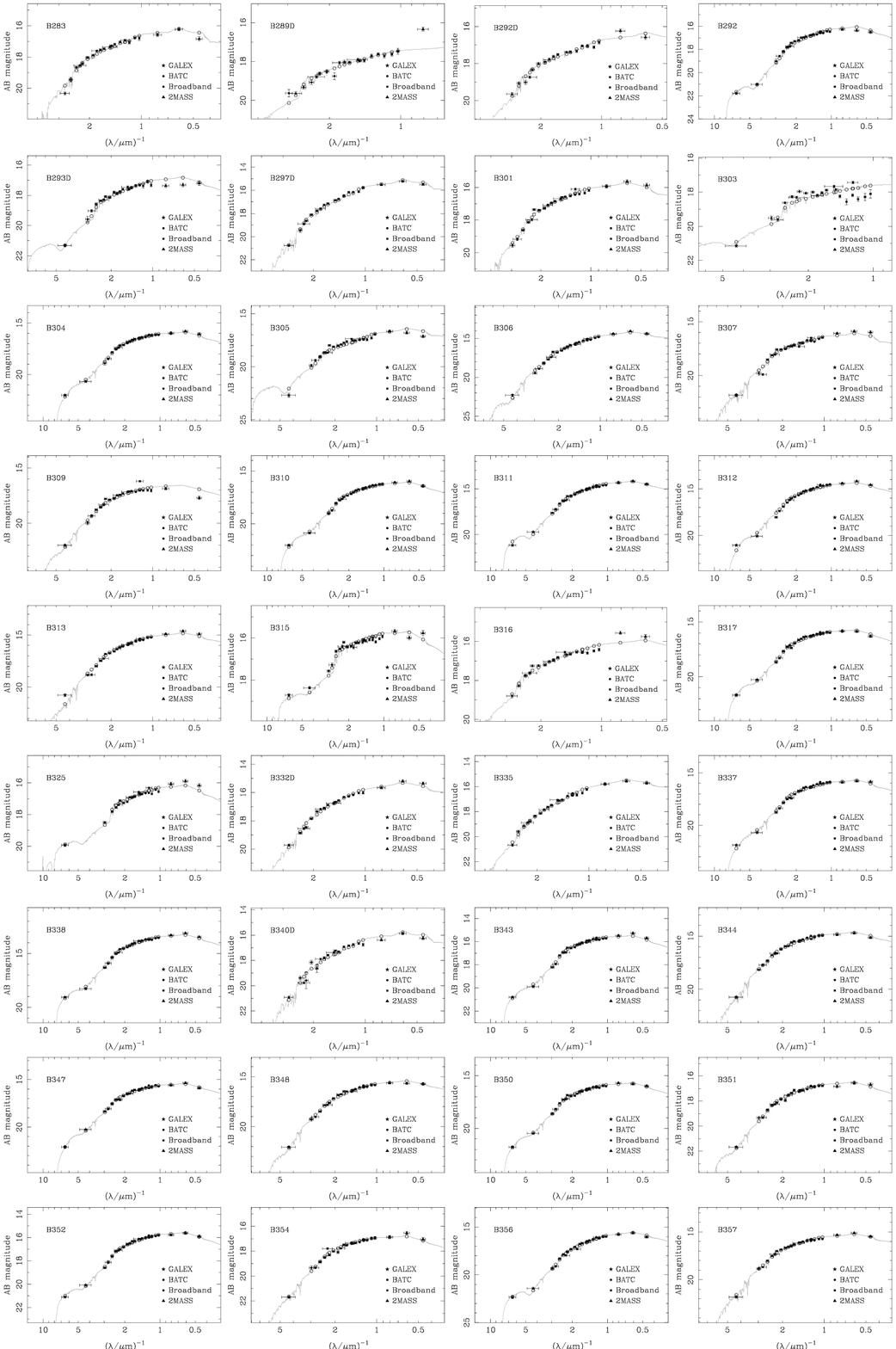}}} 
\vspace{0.2cm}
\caption{Continued.} \label{fig3}
\end{figure*}

\begin{figure*}
\vspace{-1.0cm}
\figurenum{3} \resizebox{\hsize}{!}{\rotatebox{-0}{\includegraphics{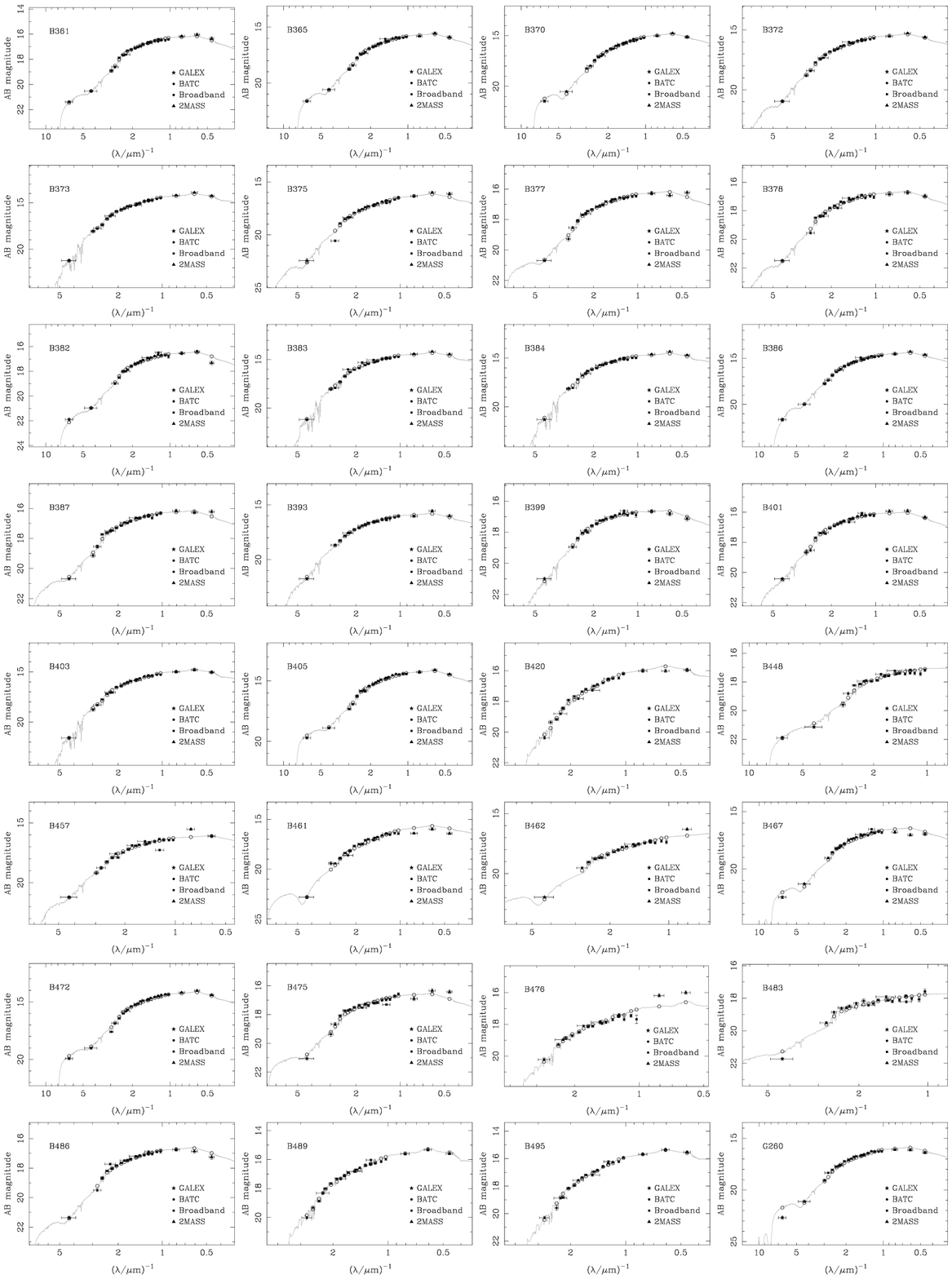}}} 
\vspace{-2.8cm}
\caption{Continued.} \label{fig3}
\end{figure*}

\section{Results and summary}
\label{result.sec}

In \S3 we determined the ages of 104 GCs and GC candidates in M31. The results are tabulated in
Table 5.  Figure 4 shows the age distribution of the sample clusters, from which we conclude that,
except for 20 clusters, the ages of most sample GCs are between 1 and 5 Gyr, with a peak at $\sim
2$ Gyr. The `usual' complement of well-known old GCs (i.e., GCs of similar age as the majority of
the Galactic GCs) is also present. In addition, while fitting SSP models to the observed data, we
found that some photometric data of a small number of clusters cannot be fitted with any SSP
models. We therefore did not use these deviating photometric data points to obtain the best fits.
This applies to the {\sl GALEX} far-UV data of B138D, the 2MASS $K_{\rm s}, H$, and $J$ magnitudes
of B142D, B181D, and B289D, respectively, the $B$-band and 2MASS $H$ fluxes of B245D, and the
$V$-band magnitude of B257.

Other authors have also considered the age distribution of the GCs in M31. For example,
\citet{bh00} discovered that M31 contains GCs exhibiting strong Balmer lines and A-type spectra,
from which one infers that these objects must be very young. \citet{Beasley04} and \citet{puzia05}
confirmed this conclusion. \citet{Burstein04} and \citet{Fusi05} carefully studied the sample of
young M31 GCs. Very recently, \citet{caldwell09} determined the ages and reddening values of 140
young clusters in M31 by comparing the observed spectra with models, and found that these clusters
are less than 2 Gyr old, while most clusters have ages between $10^8$ and $10^9$ yr.
\citet{perina09} estimated an age for VDB0-B195D of $\sim 25$~Myr based on {\sl HST}/WFPC2
color-magnitude diagrams (CMDs). The ages of the M31 clusters determined in this paper are in
general agreement with previous determinations, which we will show in more detail below on the
basis of comparisons between our determinations and previous age estimates for individual objects.

The most direct and most accurate method to determine a cluster's age is by means of main-sequence
photometry, since the absolute magnitude of the main-sequence turnoff is a strong function of age.
\citet{Williams01a,Williams01b} estimated ages of many young disk clusters in M31 based on {\sl
HST}/WFPC2 CMDs and isochrone fitting to either the main sequence or luminous evolved stars. Only
one of their clusters (B315) is in common with our sample. They obtained an age of $\sim 0.1$ Gyr
for this cluster, while we determined it to be approximately 0.5 Gyr old. Both age determinations
are mutually consistent within the uncertainties. \citet{caldwell09} compared their ages with
those of \citet{Williams01a,Williams01b} and concluded that both sets of age determinations were
in good agreement. We therefore compare our ages with \citet{caldwell09} for the seven clusters we
have in common with their sample (B018, B307, B316, B448, B475, B483, and V031: see Table 6). It
is evident that they are largely internally consistent. \citet{puzia05} also presented
spectroscopic ages, metallicities, and ${\rm [\alpha/Fe]}$ ratios for 70 M31 GCs based on Lick
line-index measurements. A cross correlation with \citet{puzia05}'s sample shows that we have 21
clusters in common. A direct comparison shows that the ages of \citet{puzia05} are systematically
older than ours. This surprising result prompted us to compare the ages of clusters in common
between \citet{puzia05} and other authors \citep{Williams01a,Williams01b,Beasley04,caldwell09}. We
found similar systematic offsets \citep[see for details also][]{ma09}.

\begin{figure}
\figurenum{4} \resizebox{\hsize}{!}{\rotatebox{-90}{\includegraphics{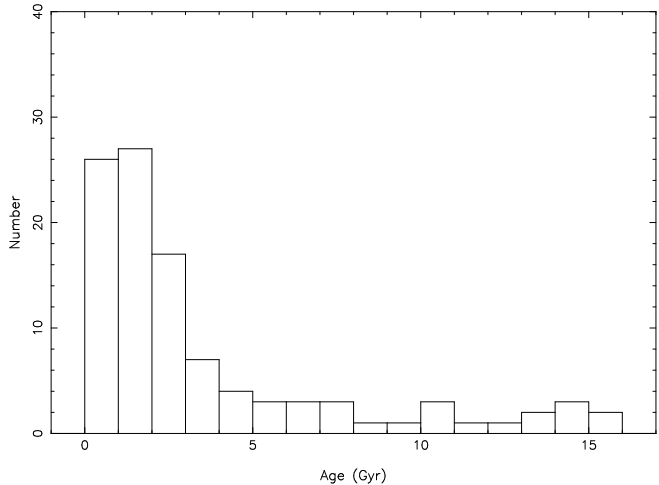}}}
\caption{Age distribution of our sample GCs and GC candidates in
 M31.} \label{fig4}
\end{figure}

We have determined the ages of M31 GCs and GC candidates in a series of previous papers
\citep{jiang03,ma06,ma06b,fan06,ma07a,ma09,ma09b} based on the same method as used in the present
paper, i.e., by constructing SEDs of known M31 GC candidates and using the SED shapes to estimate
cluster ages. In the first paper of this series, \citet{jiang03} estimated the ages of 172 M31 GC
candidates based on photometric measurements in 13 BATC intermediate-band filters and the SSP
models of Bruzual \& Charlot (1996; unpublished, hereafter BC96). Subsequently, \citet{fan06}
obtained new age estimates for 91 GCs from the \citet{jiang03} sample, based on improved
photometric data including intermediate- and broad-band magnitudes from the optical to the NIR,
and the SSP models of BC03. \citet{ma06b} then estimated the ages of 33 M31 GC candidates using
photometry in 13 BATC intermediate-band filters and the BC03 SSP models, while \citet{ma09}
determined the ages of 35 M31 GCs and GC candidates based on photometry including far- and near-UV
{\sl GALEX} observations, $UBVRI$, 13 BATC intermediate-band filters, and 2MASS $JHK_{\rm s}$,
combined with the {\sc galev} SSP models. \citet{ma06,ma07a,ma09b} determined the ages of three
specific M31 GCs (037-B327, S312, and G1) based on the BC03 SSP models and a large number of
photometric measurements. We determined the ages of these three M31 GCs for special reasons: S312
is among the first extragalactic GCs whose age was estimated accurately using main-sequence
photometry, while 037-B327 and G1 are among the most massive GCs in the Local Group.  They have
been speculated to be nucleated dwarf galaxies instead of genuine GCs \citep[see for detailed
discusions][]{ma06c,ma07b}. In this series of seven articles, we published ages for 331 different
M31 GCs and GC candidates. Figure 5 show the age distribution of these 331 objects. We see that
$\sim 40$ clusters are younger than 1 Gyr. The ages range from $< 1$ to 20 Gyr (the upper age
limit in the BC96 and BC03 SSP models). A population of young clusters, peaking at $\sim 3$ Gyr,
is also apparent.

\begin{figure}
\figurenum{5} \resizebox{\hsize}{!}{\rotatebox{-90}{\includegraphics{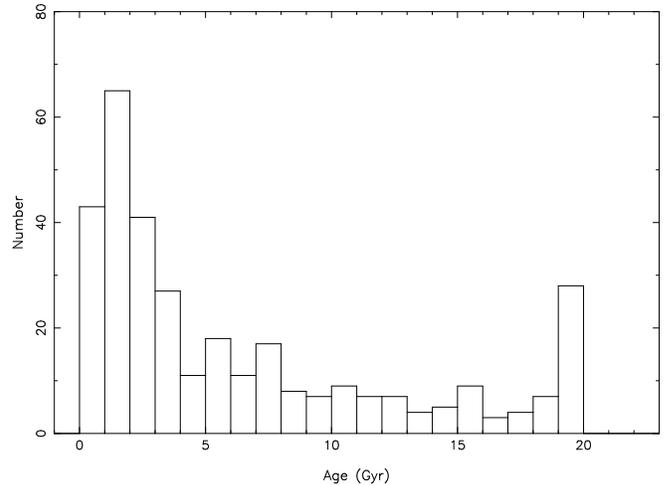}}}
\caption{Homogenized age distribution of the 331 M31 GCs and GC
candidates discussed in our series of papers.} \label{fig5}
\end{figure}

Figure 6 shows the absolute magnitudes of our sample of M31 GCs and GC candidates as a function of
their age. The crosses indicate that the ages are from \citet{jiang03},
\citet{ma06,ma06b,ma07a,ma09}, and \citet{fan06}, which were obtained based on the SSP models of
BC96 or BC03, while the circles mean that the ages are from \citet{ma09} and the present paper,
obtained on the basis of the {\sc galev} SSP models. The dashed and solid lines represent SSP
models with $Z=0.004$ taken from BC03 and {\sc galev}, respectively, for masses of $10^2$, $10^3$,
$10^4$, $10^5$, and $10^6 M_\odot$ and assuming a Salpeter stellar IMF. The $V$-band photometry is
from the RBC v.3.5. The absolute magnitudes have been corrected for extinction \citep{bh00,
fan08}, except for 037-B327, S312, and G1, the reddening values of which are from \citet{ma06},
\citet{ma07a}, and \citet{ma09b}, respectively. We adopt a distance modulus of $(m-M)_{0}=24.47$
mag \citep{McConnachie05}. Figure 6 shows that the majority of the clusters have masses between
$10^3$ and $10^6 M_\odot$.

\begin{figure}
\figurenum{6} \resizebox{\hsize}{!}{\rotatebox{-90}{\includegraphics{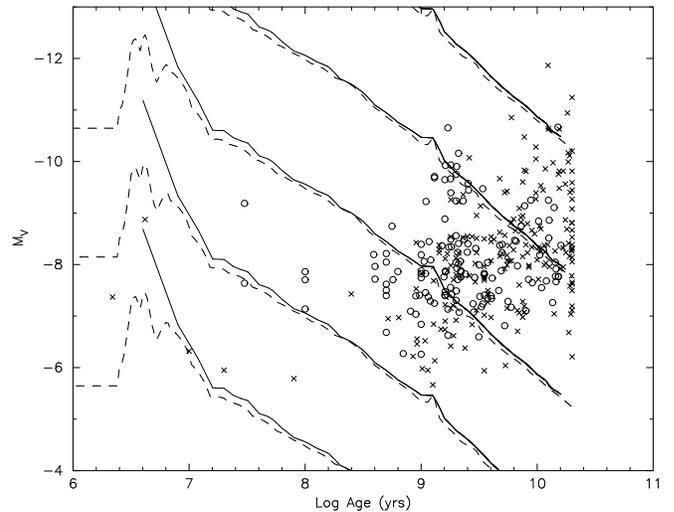}}} \caption{Absolute
$V$-band magnitudes for the M31 GCs and GC candidates as a function of age. Overplotted are
theoretical lines corresponding to (from bottom to top) masses of $10^2$, $10^3$, $10^4$, $10^5$
and $10^6 M_\odot$ from BC03 (dashed line) and {\sc galev} (solid line), respectively.}
\label{fig6}
\end{figure}

The distribution of absolute $V$ magnitude of GCs in M31 is shown in Figure 7. Overall, the
distribution has a cutoff at the faint end with a magnitude limit of about $-5.5$ mag (with a few
fainter clusters still visible, probably because of advantageous positions, e.g., observable
through a hole in the extinction distribution). The various cluster ages are separated in Figure
8, which are: (i) very young ($t< 1$ Gyr), (ii) young ($1 \leq t< 4$ Gyr), (iii) intermediate-age
($4 \leq t< 10$ Gyr), and (iv) old GCs and GC candidates ($t\geq 10$ Gyr). We do not see a clear
trend between age and brightness. However, the youngest clusters are not the most massive objects,
implying that the conditions in the M31 have not been conducive to massive cluster formation in
the recent past.

\begin{figure}
\figurenum{7} \resizebox{\hsize}{!}{\rotatebox{-90}{\includegraphics{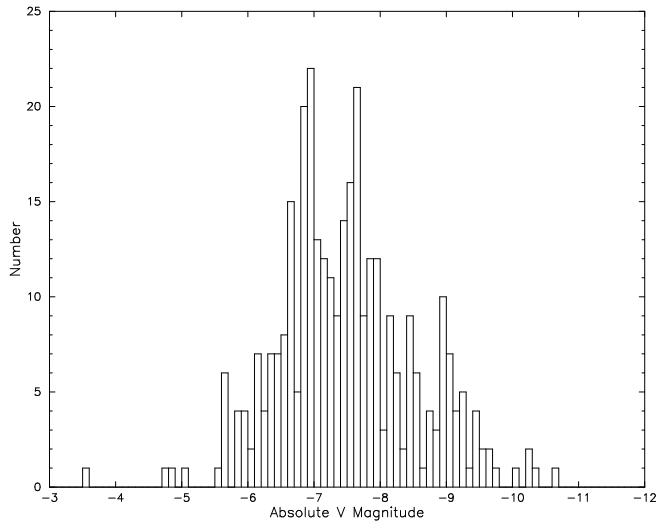}}} \caption{Histogram
of the absolute $V$ magnitude for the 331 sample GCs and GC candidates in M31.} \label{fig7}
\end{figure}

\begin{figure}
\figurenum{8} \resizebox{\hsize}{!}{\rotatebox{-90}{\includegraphics{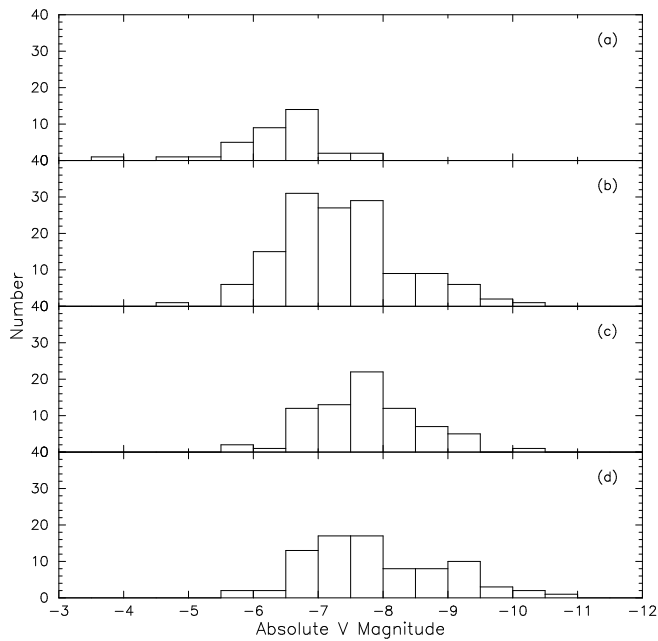}}} \caption{Histogram
of the absolute $V$ magnitude for the M31 GCs and GC candidates: (a) very young ($t< 1$ Gyr), (b)
young ($1 \leq t< 4$ Gyr), (c) intermediate-age ($4 \leq t< 10$ Gyr), and (d) old GCs and GC
candidates ($t\geq 10$ Gyr).} \label{fig8}
\end{figure}

We converted the absolute magnitudes of our M31 GC sample to photometric masses using the
appropriate age-dependent mass-to-light ratios provided by the BC03 and {\sc galev} SSP models.
The GC mass versus age diagram is shown in Figure 9. The crosses indicate that the ages are from
\citet{jiang03}, \citet{ma06,ma06b,ma07a,ma09}, and \citet{fan06}, and the masses were obtained
based on the SSP models of BC03, while the circles mean that the ages are from \citet{ma09} and
the present paper, and the masses were obtained on the basis of the {\sc galev} SSP models.
Overplotted is the fading limit, assuming $M_{V,{\rm limit}} = -5.5$ mag and evolutionary fading
based on the $Z=0.004$ BC03 (dashed line) and {\sc galev} (solid line) models, assuming a Salpeter
stellar IMF. Figure 9 shows that our observational ($\sim 50$\%) completeness limit describes the
lower mass limit of the entire GC sample up to the oldest ages very well. Similarly, the upper
envelope of the points in Figure 9 is likely a result of the `size-of sample effect'
\citep[e.g.,][and references therein]{gb08}. It is clear, however, that massive star cluster
formation halted abruptly in the disk of M31 approximately 1 Gyr ago. Given that massive ($> 10^4
M_\odot$) young ($< 1$ Gyr-old) clusters will be significantly brighter than the much older
GC-type counterparts in M31, we would have expected any such young massive clusters to have been
detected in M31, yet they have not.

\begin{figure}
\figurenum{9} \resizebox{\hsize}{!}{\rotatebox{-90}{\includegraphics{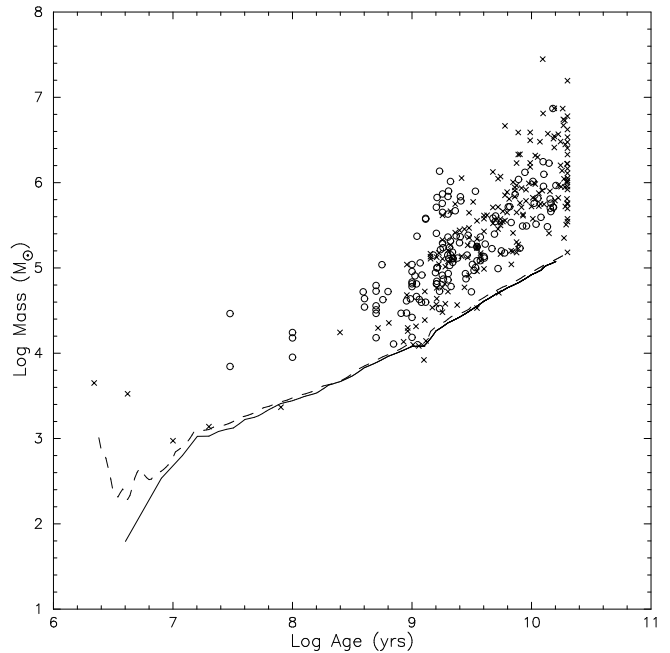}}}
\caption{Distribution of the M31 GCs and GC candidates in the age-versus-mass plane. Overplotted
is the fading limit, based on the observed $M_V = -5.5$ mag sample cutoff and the fading function
from the cluster evolutionary models with $Z=0.004$ taken from BC03 (dashed line) and {\sc galev}
(solid line), assuming a Salpeter stellar IMF.} \label{fig9}
\end{figure}

Using these 331 GCs and GC candidates with homogeneously determined ages, we can now investigate
their spatial distribution. We use an $X,Y$ projection to refer to the relative positions of the
objects. Our adopted $X$ coordinate projects along M31's major axis, where positive $X$ increases
towards the northeast, while the $Y$ coordinate extends along the minor axis of the M31 disk,
increasing towards the northwest. To obtain the relative coordinates of the M31 clusters, we
adopted $\alpha_0=00^{\rm h}42^{\rm m}44^{\rm s}.30$ and $\delta_0=+41^\circ16'09''.0$ (J2000.0)
for M31's center, following \citet{hbk91} and \citet{per02}. Formally,
\begin{equation}
X=A\sin\theta+B\cos\theta \quad {\rm and}
\end{equation}

\begin{equation}
Y=-A\cos\theta+B\sin\theta ,
\end{equation}
where $A=\sin(\alpha-\alpha_0)\cos\delta$ and $B=\sin\delta \cos\delta_0 - \cos(\alpha-\alpha_0)
\cos\delta \sin\delta_0$. We adopt a position angle of $\theta=38^\circ$ for the major axis of M31
\citep{ken89}. We divided the GCs and GC candidates into four age groups: (i) very young ($t< 1$
Gyr), (ii) young ($1 \leq t< 4$ Gyr), (iii) intermediate-age ($4 \leq t< 10$ Gyr), and (iv) old
GCs and GC candidates ($t\geq 10$ Gyr). Figure 10 shows their spatial distributions. Although our
sample of M31 GCs and GC candidates is not complete (in spatial, radial terms, given that we are
limited by the six observed fields), we note that there is a tendency for young GCs and GC
candidates to be nearly uniformly distributed around M31. The majority of old GCs appear to occupy
the central regions of the galaxy, although this restricted distribution may be caused by
selection biases. Figure 11 shows the number of GCs and GC candidates as a function of projected
radial distance from the M31 center, confirming our conclusions derived from Figure 10. Figure 12
displays the cluster ages as a function of projected radial distance. The crosses indicate that
the ages are from \citet{jiang03}, \citet{ma06,ma06b,ma07a,ma09b}, and \citet{fan06}, which were
obtained using the BC96 or BC03 SSP models, while the circles indicate that the ages are from
\citet{ma09} and the present paper, obtained on the basis of the {\sc galev} SSP models.  Figure
12 shows that young GCs and GC candidates are distributed nearly uniformly, and that most of the
old GCs (and candidates) are more concentrated.

\begin{figure}
\figurenum{10} \resizebox{\hsize}{!}{\rotatebox{0}{\includegraphics{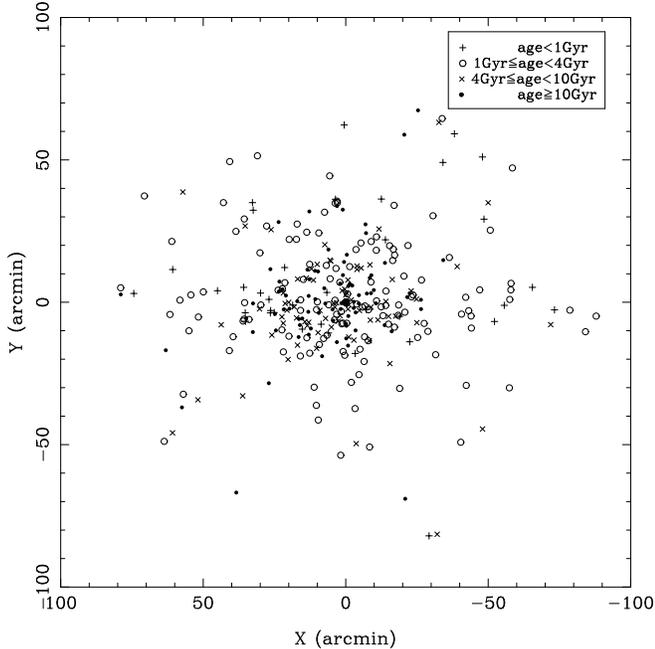}}} \caption{Spatial
distribution of the M31 GCs and GC candidates: very young ($t< 1$ Gyr), young ($1 \leq t< 4$ Gyr),
intermediate-age ($4 \leq t< 10$ Gyr) and old GCs and GC candidates ($t\geq 10$ Gyr).}
\label{fig10}
\end{figure}

\begin{figure}
\figurenum{11} \resizebox{\hsize}{!}{\rotatebox{-90}{\includegraphics{fig11.ps}}} \caption{Radial
distribution of the M31 GCs and GC candidates: (a) very young ($t< 1$ Gyr), (b) young ($1 \leq t<
4$ Gyr), (c) intermediate-age ($4 \leq t< 10$ Gyr), and (d) old GCs and GC candidates ($t\geq 10$
Gyr).} \label{fig11}
\end{figure}

\begin{figure}
\figurenum{12} \resizebox{\hsize}{!}{\rotatebox{-90}{\includegraphics{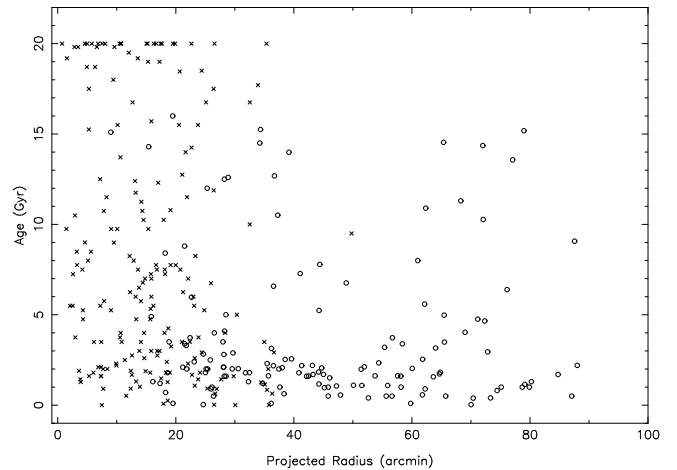}}} \caption{Age
versus projected galactocentric radius for 331 M31 GCs and GC candidates.} \label{fig12}
\end{figure}

This paper presents photometry of 104 M31 globular clusters (GCs) and GC candidates in 15
intermediate-band filters of the BATC photometric system. The age of the clusters were obtained by
comparing the photometric data with the theoretical synthesis models. The ages of our sample
clusters cover a large range, although most clusters are younger than 10 Gyr. Combined with the
ages obtained in our series of previous papers focusing on the M31 GC system, we present the full
M31 GC age distribution. The results show that the M31 GC system contains populations of young and
intermediate-age GCs, as well as the `usual' complement of well-known old GCs, i.e., GCs of
similar age as the majority of the Galactic GCs. In addition, young GCs (and GC candidates) are
distributed nearly uniformly in radial distance from the center of M31, while most old GCs (and GC
candidates) are more strongly concentrated.

\newpage
\acknowledgments {We are indebted to the referee for thoughtful
comments and insightful suggestions that improved this paper
significantly. This work was supported by the Chinese National Natural
Science Foundation (grants 10873016, 10633020, 10603006, and 10803007)
and by National Basic Research Program of China (973 Program; grant
2007CB815403).}

\begin{table*}
\begin{center}
\caption{BATC filter parameters}
\label{t1.tab}
\begin{tabular}{ccc}
\tableline\tableline Filter & Central wavelength & Bandwidth \\
 &  (\AA) & (\AA)\\
\tableline
 $a$ & 3360 & 360 \\
 $b$ & 3890 & 340 \\
 $c$ & 4210 & 320 \\
 $d$ & 4550 & 340 \\
 $e$ & 4925 & 390 \\
 $f$ & 5270 & 340 \\
 $g$ & 5795 & 310 \\
 $h$ & 6075 & 310 \\
 $i$ & 6660 & 480 \\
 $j$ & 7050 & 300 \\
 $k$ & 7490 & 330 \\
 $m$ & 8020 & 260 \\
 $n$ & 8480 & 180 \\
 $o$ & 9190 & 260 \\
 $p$ & 9745 & 270 \\
\tableline
\end{tabular}\\
\end{center}
\end{table*}

\begin{table*}
\begin{center}
\caption{BATC intermediate-band photometry (magnitudes) of 104 M31 GCs and GC
  candidates.} 
\label{t2.tab}
\begin{tabular}{cccccccccccccccc}
\tableline\tableline
Object & $a$& $b$& $c$& $d$& $e$& $f$& $g$& $h$& $i$& $j$&$ k$& $m$& $n$& $o$& $p$ \\
\hline
 B001   &19.08  &18.94  &18.32  &17.68  &17.49  &17.23  &16.78  &16.65  &16.37  &16.28  &16.10  &15.89  &15.89  &15.72  &15.67  \\
        &0.190  &0.045  &0.022  &0.038  &0.032  &0.016  &0.021  &0.011  &0.008  &0.013  &0.012  &0.011  &0.016  &0.021  &0.019  \\
 B003   &...    &18.67  &18.35  &18.01  &17.93  &17.74  &17.38  &17.28  &17.15  &17.10  &16.98  &16.85  &16.80  &16.77  &16.69  \\
        &...    &0.039  &0.022  &0.045  &0.046  &0.023  &0.031  &0.019  &0.015  &0.024  &0.024  &0.021  &0.043  &0.044  &0.051  \\
 B005   &18.07  &17.23  &16.66  &16.01  &15.94  &15.78  &15.43  &15.32  &15.08  &15.05  &14.90  &14.86  &14.86  &14.68  &14.62  \\
        &0.037  &0.019  &0.017  &0.016  &0.012  &0.010  &0.010  &0.008  &0.006  &0.007  &0.009  &0.007  &0.008  &0.011  &0.013  \\
 B018   &19.68  &18.58  &18.18  &17.84  &17.77  &17.69  &17.36  &17.31  &17.12  &17.07  &16.89  &16.90  &16.94  &16.82  &16.87  \\
\end{tabular}
\end{center}
\end{table*}
\addtocounter{table}{-1}

\begin{table*}
\begin{center}
\caption{Continued.} 
\label{t2.tab}
\begin{tabular}{cccccccccccccccc}
\tableline\tableline
Object & $a$ &$b$ &$c$ & $d$& $e$ &$f$ &$g$ &$h$ &$i$ &$j$ &$k$ &$m$ &$n$ &$o$ & $p$ \\
\hline
 B266   &...    &...    &20.00  &19.47  &18.86  &18.67  &18.05  &17.93  &17.63  &17.49  &17.30  &17.14  &17.02  &16.91  &16.83  \\
        &...    &...    &0.150  &0.103  &0.065  &0.056  &0.042  &0.037  &0.029  &0.031  &0.030  &0.031  &0.052  &0.038  &0.059  \\
 B270D  &...    &18.62  &18.25  &18.05  &17.84  &17.62  &17.42  &17.27  &17.15  &17.05  &16.98  &16.92  &16.87  &16.93  &16.83  \\
        &...    &0.052  &0.020  &0.043  &0.026  &0.019  &0.030  &0.012  &0.012  &0.018  &0.019  &0.017  &0.031  &0.033  &0.129  \\
 B272   &...    &...    &19.75  &19.01  &18.61  &18.39  &17.91  &17.82  &17.58  &17.40  &17.16  &17.02  &16.95  &16.74  &16.56  \\
        &...    &...    &0.084  &0.054  &0.039  &0.031  &0.033  &0.028  &0.026  &0.026  &0.035  &0.027  &0.056  &0.034  &0.046  \\
 B281   &...    &...    &18.62  &18.23  &17.99  &17.80  &17.53  &17.42  &17.23  &17.16  &17.00  &16.90  &16.90  &16.67  &16.67  \\
\end{tabular}
\end{center}
\end{table*}
\addtocounter{table}{-1}

\begin{table*}
\begin{center}
\caption{Continued.} 
\label{t2.tab}
\begin{tabular}{cccccccccccccccc}
\tableline\tableline
Object & $a$ &$b$ &$c$ & $d$& $e$ &$f$ &$g$ &$h$ &$i$ &$j$ &$k$ &$m$ &$n$ &$o$ & $p$ \\
\hline
 B343   &18.22  &17.19  &16.94  &16.53  &16.55  &16.48  &16.20  &16.11  &15.91  &15.92  &15.79  &15.80  &15.84  &15.74  &15.72  \\
        &0.039  &0.012  &0.016  &0.019  &0.012  &0.010  &0.012  &0.008  &0.005  &0.009  &0.012  &0.008  &0.013  &0.017  &0.028  \\
 B344   &18.17  &17.32  &16.79  &16.62  &16.22  &15.96  &15.59  &15.49  &15.40  &15.32  &15.16  &14.98  &15.07  &14.95  &14.93  \\
        &0.063  &0.015  &0.009  &0.018  &0.015  &0.008  &0.012  &0.006  &0.006  &0.010  &0.007  &0.006  &0.008  &0.015  &0.015  \\
 B347   &18.35  &17.54  &17.13  &17.15  &16.71  &16.51  &16.17  &16.09  &16.03  &15.94  &15.84  &15.63  &15.75  &15.60  &15.71  \\
        &0.064  &0.017  &0.011  &0.026  &0.023  &0.012  &0.018  &0.009  &0.011  &0.016  &0.011  &0.010  &0.020  &0.030  &0.028  \\
 B348   &19.28  &18.49  &17.86  &17.63  &17.18  &16.84  &16.47  &16.48  &16.36  &16.26  &16.08  &15.89  &15.98  &15.83  &15.84  \\
\end{tabular}
\end{center}
\end{table*}
\addtocounter{table}{-1}

\begin{table*}
\begin{center}
\caption{Continued.} 
\label{t2.tab}
\begin{tabular}{cccccccccccccccc}
\tableline\tableline
Object & $a$ &$b$ &$c$ & $d$& $e$ &$f$ &$g$ &$h$ &$i$ &$j$ &$k$ &$m$ &$n$ &$o$ & $p$ \\
\hline
 B461   &19.41  &18.91  &18.38  &18.15  &17.67  &17.49  &17.09  &16.99  &16.85  &16.78  &16.62  &16.41  &16.54  &16.39  &16.43  \\
        &0.157  &0.056  &0.027  &0.067  &0.047  &0.020  &0.036  &0.016  &0.018  &0.034  &0.017  &0.014  &0.026  &0.050  &0.054  \\
 B462   &...    &19.03  &18.77  &18.67  &18.42  &18.19  &17.80  &17.74  &17.66  &17.52  &17.45  &17.30  &17.44  &17.29  &17.37  \\
        &...    &0.063  &0.040  &0.112  &0.092  &0.036  &0.065  &0.030  &0.036  &0.061  &0.032  &0.027  &0.058  &0.090  &0.130  \\
 B467   &...    &18.57  &18.25  &18.00  &17.83  &17.51  &17.35  &17.25  &17.11  &17.02  &17.08  &16.82  &16.86  &16.75  &16.84  \\
        &...    &0.052  &0.052  &0.044  &0.035  &0.060  &0.039  &0.021  &0.012  &0.024  &0.038  &0.019  &0.032  &0.046  &0.063  \\
 B472   &17.60  &16.43  &16.00  &15.68  &15.40  &15.25  &14.95  &14.89  &14.75  &14.66  &14.58  &14.46  &14.40  &14.35  &14.35  \\
\end{tabular}
\end{center}
\end{table*}

\begin{table*}
\begin{center}
\caption{{\sl GALEX}, optical broad-band, and 2MASS NIR photometry of
  104 M31 GCs and GC candidates.}
\label{t3.tab}
\begin{tabular}{cccccccccccc}
\tableline\tableline
Object & $c$\dag  &  FUV & NUV & $U$ & $B$ & $V$ & $R$ & $I$ & $J$ & $H$  & $K_{\rm s}$ \\
\hline
 B001  &1     &...   &...   &18.82 &18.33 &17.06 &16.47 &15.41 &14.69 &13.73 &13.84 \\
         &&...   &...   &0.08  &0.05  &0.05  &0.05  &0.05  &0.04  &0.04  &0.05  \\
 B003  &1     &23.06 &21.55 &18.40 &18.35 &17.57 &17.07 &16.41 &15.95 &15.15 &15.51 \\
         &&0.20  &0.04  &0.04  &0.02  &0.01  &0.03  &0.02  &0.08  &0.10  &0.12  \\
 B005  &1     &...   &21.08 &16.12 &16.04 &15.44 &14.99 &14.66 &13.40 &12.69 &12.51 \\
         &&...   &0.12  &0.02  &0.01  &0.01  &0.01  &0.01  &0.03  &0.03  &0.03  \\
 B018  &1     &...   &22.09 &18.47 &18.25 &17.53 &17.00 &16.38 &15.47 &14.77 &14.59 \\
\end{tabular}\\
\tablenotetext{\dag}{New classification flag (RBC v.3.5):
 1 = GC, 2 = candidate GC.}
\end{center}
\end{table*}
\addtocounter{table}{-1}

\begin{table*}
\begin{center}
\caption{Continued.} \label{t3.tab}
\begin{tabular}{cccccccccccc}
\tableline\tableline
Object & $c$\dag  &  FUV & NUV & $U$ & $B$ & $V$ & $R$ & $I$ & $J$ & $H$  & $K_{\rm s}$ \\
\hline
 B261D &2     &...   &21.31 &18.02 &18.06 &17.60 &16.97 &...   &15.35 &15.38 &...   \\
         &&...   &0.04  &0.08  &0.05  &0.05  &0.05  &...   &0.08  &0.10  &...   \\
 B266  &1     &...   &...   &18.76 &19.29 &18.32 &17.40 &16.50 &15.64 &15.06 &14.45 \\
         &&...   &...   &0.08  &0.05  &0.05  &0.05  &0.05  &0.08  &0.10  &0.10  \\
 B270D &2     &...   &21.81 &18.24 &17.90 &17.50 &16.94 &...   &15.43 &15.09 &...   \\
         &&...   &0.07  &0.08  &0.05  &0.05  &0.05  &...   &0.08  &0.10  &...   \\
 B272  &1     &...   &...   &19.94 &19.52 &18.20 &17.50 &16.61 &14.70 &14.40 &...   \\
\end{tabular}\\
\tablenotetext{\dag}{New classification flag (RBC v.3.5):
 1 = GC, 2 = candidate GC.}
\end{center}
\end{table*}
\addtocounter{table}{-1}

\begin{table*}
\begin{center}
\caption{Continued.} \label{t3.tab}
\begin{tabular}{cccccccccccc}
\tableline\tableline
Object & $c$\dag  &  FUV & NUV & $U$ & $B$ & $V$ & $R$ & $I$ & $J$ & $H$  & $K_{\rm s}$ \\
\hline
 B338  &1     &19.05 &18.28 &15.21 &15.06 &14.30 &13.67 &13.28 &12.45 &11.79 &11.67 \\
         &&0.02  &0.01  &0.01  &0.01  &0.01  &0.05  &0.01  &0.02  &0.02  &0.02  \\
 B340D &2     &...   &...   &20.22 &19.93 &17.92 &17.22 &...   &15.53 &14.55 &14.45 \\
         &&...   &...   &0.08  &0.05  &0.05  &0.05  &...   &0.08  &0.07  &0.10  \\
 B343  &1     &20.80 &19.90 &17.23 &17.08 &16.31 &15.91 &15.27 &14.63 &13.94 &13.90 \\
         &&0.06  &0.02  &0.08  &0.05  &0.05  &0.05  &0.05  &0.04  &0.04  &0.05  \\
 B344  &1     &...   &20.75 &17.01 &16.77 &15.95 &15.33 &14.87 &14.02 &13.40 &13.35 \\
\end{tabular}\\
\tablenotetext{\dag}{New classification flag (RBC v.3.5):
 1 = GC, 2 = candidate GC.}
\end{center}
\end{table*}
\addtocounter{table}{-1}

\begin{table*}
\begin{center}
\caption{Continued.} \label{t3.tab}
\begin{tabular}{cccccccccccc}
\tableline\tableline
Object & $c$\dag  &  FUV & NUV & $U$ & $B$ & $V$ & $R$ & $I$ & $J$ & $H$  & $K_{\rm s}$ \\
\hline
 B420  &2     &...   &...   &19.68 &18.96 &17.85 &17.12 &16.13 &15.16 &14.68 &14.15 \\
         &&...   &...   &0.08  &0.05  &0.05  &0.05  &0.05  &0.08  &0.07  &0.10  \\
 B448  &1     &21.90 &21.14 &18.11 &18.10 &17.49 &17.06 &16.77 &...   &...   &...   \\
         &&0.12  &0.05  &0.09  &0.04  &0.01  &0.04  &0.03  &...   &...   &...   \\
 B457  &1     &...   &21.21 &18.05 &17.73 &16.91 &16.38 &16.87 &14.68 &14.78 &...   \\
         &&...   &0.03  &0.08  &0.05  &0.05  &0.05  &0.05  &0.04  &0.07  &...   \\
 B461  &1     &...   &22.82 &18.73 &18.77 &17.52 &16.89 &16.04 &15.55 &14.67 &14.61 \\
\end{tabular}\\
\tablenotetext{\dag}{New classification flag (RBC v.3.5):
 1 = GC, 2 = candidate GC.}
\end{center}
\end{table*}

\begin{table}
\begin{center}
\caption{Reddening values (magnitudes) and metallicities (dex) for 104 M31 GCs and GC candidates.} \label{t4.tab}
\begin{tabular}{ccccc}
\tableline \tableline
Object & $E(B-V)$& ref.$^a$ & \multicolumn{1}{c}{$\rm [Fe/H]$}& ref.$^b$\\
\hline
 B001  &   0.25$\pm$   0.02&  1&  $-0.58\pm$   0.18&  1\\
 B003  &   0.19$\pm$   0.02&  1&  $-2.08\pm$   0.07&  4\\
 B005  &   0.28$\pm$   0.02&  1&  $-1.18\pm$   0.17&  1\\
 B018  &   0.20$\pm$   0.01&  1&  $-1.63\pm$   0.77&  1\\
 B020D &   0.22$\pm$   0.06&  1&  $-0.76\pm$   0.08&  4\\
 B024  &   0.03$\pm$   0.02&  1&  $-0.48\pm$   0.30&  3\\
 B046  &   0.19$\pm$   0.03&  1&  $-1.84\pm$   0.61&  3\\
\end{tabular}\\
\end{center}
\tablenotetext{a}{The reddening values are taken from
\citet{fan08} (ref=1) and \citet{bh00} (ref=2).}
\tablenotetext{b}{The metallicities are taken from \citet{per02}
(ref=1), \citet{bh00} (ref=2), \citet{hbk91} (ref=3), and \citet{fan08}
(ref=4).}
\end{table}
\addtocounter{table}{-1}

\begin{table}
\begin{center}
\caption{Continuted.} \label{t4.tab} 
\begin{tabular}{ccccc}
\tableline \tableline
Object & $E(B-V)$& ref.$^a$ & \multicolumn{1}{c}{$\rm [Fe/H]$}& ref.$^b$\\
\hline
 B313  &   0.21$\pm$   0.02&  1&  $-1.09\pm$   0.10&  1\\
 B315  &   0.07$\pm$   0.02&  1&  $-2.35\pm$   0.54&  1\\
 B316  &   0.21$\pm$   0.03&  1&  $-1.47\pm$   0.23&  1\\
 B317  &   0.11$\pm$   0.02&  1&  $-2.12\pm$   0.36&  3\\
 B325  &   0.14$\pm$   0.02&  1&  $-1.77\pm$   0.08&  4\\
 B332D &   0.33$\pm$   0.13&  1&  $-0.65\pm$   0.09&  4\\
 B335  &   0.65$\pm$   0.02&  1&  $-1.05\pm$   0.26&  1\\
\end{tabular}\\
\end{center}
\tablenotetext{a}{The reddening values are taken from
\citet{fan08} (ref=1) and \citet{bh00} (ref=2).}
\tablenotetext{b}{The metallicities are taken from \citet{per02}
(ref=1), \citet{bh00} (ref=2), \citet{hbk91} (ref=3), and \citet{fan08}
(ref=4).}
\end{table}

\begin{table*}
\begin{center}
\caption{Ages estimates for 104 GCs and GC candidates in M31.} \label{t5.tab} 
\begin{tabular}{ccc|ccc}
\tableline \tableline
Object &   Age &     $\chi_{\rm min}^2$ & Object &   Age &     $\chi_{\rm min}^2$\\
  & (Gyr) & (per degree of freedom)&  & (Gyr) & (per degree of freedom) \\
\hline
 B001    & $ 10.51  \pm  0.80  $ &     2.79 & B313    & $  7.28  \pm  0.70  $ &    5.48\\
 B003    & $  2.01  \pm  0.20  $ &     2.10 & B315    & $  0.50  \pm  0.10  $ &    7.88\\
 B005    & $  1.60  \pm  0.10  $ &    18.92 & B316    & $  1.06  \pm  0.10  $ &    7.81\\
 B018    & $  1.79  \pm  0.20  $ &     2.74 & B317    & $  2.06  \pm  0.15  $ &    2.59\\
 B020D   & $  8.41  \pm  1.70  $ &     1.29 & B325    & $  0.40  \pm  0.10  $ &    5.28\\
 B024    & $ 15.25  \pm  0.75  $ &     2.96 & B332D   & $ 11.30  \pm  0.85  $ &    4.33\\
\end{tabular}
\end{center}
\end{table*}

\begin{table*}
\begin{center}
\caption{Age comparison} \label{t6.tab} 
\begin{tabular}{ccc}
\tableline \tableline
Object &  Age (Gyr) &   Age (Gyr)  \\
     &  \citep{caldwell09}    & (this paper) \\
\hline
 B018           &1.00           &$1.79           \pm0.20           $\\
 B303           &0.40           &$0.50           \pm0.10           $\\
 B307           &1.00           &$1.61           \pm0.10           $\\
 B315           &0.16           &$0.50           \pm0.10           $\\
 B316           &1.00           &$1.06           \pm0.10           $\\
 B325           &0.63           &$0.40           \pm0.10           $\\
 B448           &0.25           &$1.70           \pm0.10           $\\
 B475           &0.32           &$0.97           \pm0.10           $\\
 B483           &0.50           &$1.00           \pm0.10           $\\
\tableline
\end{tabular}
\end{center}
\end{table*}

\end{document}